\documentclass[prl,twocolumn,superscriptaddress,notitlepage]{revtex4-1}
\usepackage{makeidx}
\usepackage[utf8]{inputenc}
\usepackage{graphicx}
\usepackage{amsmath}
\usepackage{mathrsfs}
\usepackage{graphicx}
\usepackage{subfigure}
\usepackage{dcolumn}
\usepackage{bm}
\usepackage{bbm}
\usepackage{color}
\usepackage{esint}
\usepackage{lineno}
\usepackage[breaklinks,
            colorlinks,
            urlcolor=blue,
            linkcolor=blue,
            anchorcolor=blue,
            citecolor=blue]{hyperref}
\usepackage[dvipsnames]{xcolor}
\renewcommand{\section}[1]{{\par\it #1.---}\ignorespaces}

\begin{document}

\title{Smith-Purcell radiation from time grating}

\author{Juan-Feng Zhu}
\affiliation{Science, Mathematics, and Technology (SMT), Singapore University of Technology and Design (SUTD), 8 Somapah Road, Singapore 487372.}

\author{Ayan Nussupbekov}
\affiliation{
Institute of High Performance Computing, A*STAR (Agency for Science, Technology, and Research), 1 Fusionopolis Way, No. 16-16 Connexis, Singapore 138632.
}

\author{Wenjie Zhou}
\affiliation{Science, Mathematics, and Technology (SMT), Singapore University of Technology and Design (SUTD), 8 Somapah Road, Singapore 487372.}
\affiliation{Department of Electrical and Computer Engineering, National University of Singapore, 4 Engineering Drive 3, Singapore 117583.}

\author{Zicheng Song}
\affiliation{Center for Composite Materials and Structures, School of Astronautics, Harbin Institute of Technology, Harbin 150080, China.}

\author{Xuchen Wang}
\affiliation{Institute of Nanotechnology, Karlsruhe Institute of Technology, Karlsruhe, Germany}

\author{Zi-Wen Zhang}
\affiliation{Optical Communication Systems and Networks, School of Electronics, Peking University, Beijing, 100871, P. R. China}

\author{Chao-Hai Du}
\affiliation{Optical Communication Systems and Networks, School of Electronics, Peking University, Beijing, 100871, P. R. China}

\author{Ping Bai}
\affiliation{
Institute of High Performance Computing, A*STAR (Agency for Science, Technology, and Research), 1 Fusionopolis Way, No. 16-16 Connexis, Singapore 138632.
}
\author{Ching Eng Png}
\affiliation{
Institute of High Performance Computing, A*STAR (Agency for Science, Technology, and Research), 1 Fusionopolis Way, No. 16-16 Connexis, Singapore 138632.
}
\author{Cheng-Wei Qiu}
\email{chengwei.qiu@nus.edu.sg}
\affiliation{Department of Electrical and Computer Engineering, National University of Singapore, 4 Engineering Drive 3, Singapore 117583.}

\author{Lin Wu}
\email{lin\_wu@sutd.edu.sg}
\affiliation{Science, Mathematics, and Technology (SMT), Singapore University of Technology and Design (SUTD), 8 Somapah Road, Singapore 487372.}
\affiliation{
Institute of High Performance Computing, A*STAR (Agency for Science, Technology, and Research), 1 Fusionopolis Way, No. 16-16 Connexis, Singapore 138632.
}

\begin{abstract}
Smith-Purcell radiation (SPR) occurs when an electron skims above a spatial grating, but the fixed momentum compensation from the static grating imposes limitations on the emission wavelength. It has been discovered that a temporally periodic system can provide energy compensation to generate light emissions in free space. Here, we introduce temporal SPR ($t$-SPR) emerging from a time grating and propose a generalized $t$-SPR dispersion equation to predict the relationship between radiation frequency, direction, electron velocity, modulation period, and harmonic orders. Compared to conventional SPR, $t$-SPR can: 
1) Provide a versatile platform for manipulating SPR emission through temporal modulation ($e.g.$, period, amplitude, wave shape). 
2) Exhibit strong robustness to the electron-grating separation, alleviating the constraints associated with extreme electron near-field excitation. 
3) Introduce additional energy channels through temporal modulation, enhancing and amplifying emission.
\end{abstract}

\maketitle

Smith-Purcell radiation (SPR) arises when fast-moving charged particles travel above a spatially periodic grating. This phenomenon was initially elucidated by D. H. Smith and E. M. Purcell during the 1950s \cite{smith1953visible} and has since discovered a plethora of applications spanning diverse fields, including vacuum electronic devices \cite{leavitt1979orotron}, particle accelerators \cite{nanni2015terahertz},
and free-electron lasers \cite{urata1998superradiant}. 
In recent years, SPR has achieved remarkable advancements by replacing conventional gratings with metamaterials \cite{su2019manipulating,roques2023free}, enabling comprehensive control over its coherence, polarization, and radiation direction \cite{korbly2005observation,wang2016manipulating,su2019complete,kaminer2017spectrally,jing2021polarization, jing2019spiral,zhang2023chiral}. 
In particular, novel and captivating optical phenomena such as bound states in the continuum \cite{yang2018maximal} and flat-band resonances \cite{yang2023photonic} have been harnessed to significantly enhance the intensity of SPR. 
Nevertheless, till today, all development still relies on ``momentum compensation'' facilitated by spatial modulations and energy of free electron.

A special type of metamaterial, time-varying metamaterial, whose physical properties change over time,  provides additional degrees of freedom in the time domain for manipulating interactions with light waves, beyond the conventional three-dimensional space \cite{engheta2023four,fante1971transmission}. In contrast to spatial interfaces, momentum is conserved at temporal boundaries \cite{moussa2023observation} instead of energy, because external energy is required to modulate the material properties, leading to changes in the electromagnetic energy within the system. 
Time-varying metamaterials have already demonstrated exciting applications, including time reversal \cite{bacot2016time}, nonreciprocity \cite{sounas2017non,guo2019nonreciprocal}, temporal aiming \cite{pacheco2020temporal}, and even more fascinating possibilities on the horizon \cite{pacheco2020antireflection,tirole2023double,galiffi2023broadband,lee2018linear,oue2022vcerenkov}. 
In particular, photonic time crystal (PTC) can be created by implementing temporal periodic modulation in uniform materials \cite{lustig2018topological}. 
Analogous to photonic crystals, the temporal periodic modulation leads to interference in time reflections and refractions, giving rise to the creation of bands and bandgaps in momentum. 
What's particularly exciting is the revelation of the exponential energy growth within the PTC momentum gap, providing opportunities for achieving light amplification \cite{lyubarov2022amplified,wang2023metasurface}. 
It's worth noting that the periodic temporal modulation induces frequency harmonic waves rather than momentum harmonic waves, thus offering a versatile platform for light manipulation. For instance, 
the propagating wave can be coupled to near-field surface waves along the time-grating surface through frequency down-conversion \cite{galiffi2020wood}. The temporal modulation provides additional energy to allow free electrons to induce the Cherenkov radiation (CR) within the PTC, even with mismatched energy threshold \cite{dikopoltsev2022light}.

\begin{figure}[h]
\centering
\includegraphics[scale=1]{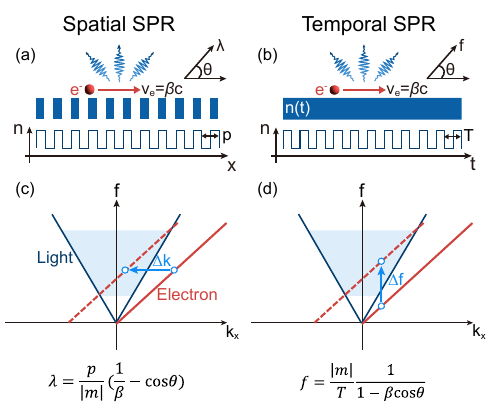}
\caption{Spatial SPR ($s$-SPR) vs. temporal SPR ($t$-SPR).
Schematics of (a) $s$-SPR or (b) $t$-SPR, induced when a swift free electron flies (with a speed of $v_e$) above (a) a grating with refractive index $n$ changing along the $x$-space with a period of $p$,
or 
(b) a time grating with $n$ changing with time $t$ at a period of $T$. 
Dispersion diagrams for (c) $s$-SPR and (d) $t$-SPR where blue lines indicate light cone. 
Diffracted by (c) the momentum harmonic wave from the space grating, or (d) the frequency harmonic wave from the TG, 
the electron dispersion  curve (red solid line) shifts (c) left/right along the $k_{x}$-axis as energy is conserved, 
or (d) upward/downward along the $f$-axis as momentum is conserved.  
The $s$-SPR and $t$-SPR are characterized by emission wavelength ($\lambda$) or frequency ($f$) and emission angle ($\theta$), which are determined by the normalized electron velocity ($\beta=v_e/c$) and modulation period ($p$ or $T$) following the shown generalized dispersion equations.
}
\label{fig1}
\end{figure}

Inspired by the work of CR in PTC \cite{dikopoltsev2022light}, we propose the concept of temporal SPR ($t$-SPR) from a time grating (TG) as illustrated in \textcolor{blue}{Fig. \ref{fig1}}, an analogy to the conventional spatial SPR ($s$-SPR) from a spatial grating.  
When the free electron flies above the TG, owing to the frequency harmonic wave, the $t$-SPR can be stimulated over a broad frequency range. 
In this Letter, we aim to extend physics concepts of conventional $s$-SPR to $t$-SPR.
We first derive the dispersion equation for $t$-SPR to reveal the relationship between the radiation frequency $f$, radiation direction $\theta$, the normalized electron velocity $\beta$, and harmonic order $m$. 
We then employ full-wave electromagnetic simulations to verify our derived dispersion equation. 
To engineer the characteristics of $t$-SPR (benchmark to $s$-SPR), we study the conditions for maximizing its radiation intensity, in terms of internal factor ($e.g.,$ electron-grating separation $d$, normalized electron velocity $\beta$) and external factor ($e.g.,$ time modulation function, period, amplitude). 
We hope to open a new research pursuit toward ``active'' free-electron light sources enabled by material temporal modulation,  beyond the state-of-the-art spatial methods.

Without losing generality, $s$-SPR is introduced first. As depicted in \textcolor{blue}{Fig. \ref{fig1}(a)}, $s$-SPR is induced when a free electron skims over a spatial grating (refractive index $n$ is modulated periodically as a function of $x$) with a velocity $v_e=\beta c$ along the $x$ direction. To illustrate the mechanism, 
the momentum-frequency dispersion diagram is analyzed in \textcolor{blue}{Fig. \ref{fig1}(c)}, where blue line indicates the light cone. The dispersion relation of the free electron can be expressed as $\omega=v_{e}k_{xe}$ (red solid line), where $k_{xe}$ is the propagation wavenumber along the $x$-axis. Since $v_{e}<c$, we have $k_{xe}>k_{0}$ with $k_{0}$ the free space wavenumber, implying that the electromagnetic wave carried by the free electron is evanescent. 
Spatial grating provides the free electron with momentum compensation: 
$\Delta k_{x} =2\pi m/p$, where $m$ denotes the diffraction order, and $p$ is the period of grating. 
Consequently, the dispersion curve for the free electron can be shifted left $(m<0)$ or right $(m>0)$ along the $k_{x}$-axis, as exemplified by the blue arrow in \textcolor{blue}{Fig. \ref{fig1}(c)}. 
When the electron dispersion curve shifts above the light cone (red dashed line), the evanescent wave can diffract into a propagating wave in free space. 
The corresponding radiation frequency range can be determined by the intersection points between the dispersion curve of the electron and the light cone (blue region). The classic dispersion relation between radiation frequency (or wavelength $\lambda=c/f$) and angle ($\theta$) can be obtained via the momentum-matching conditions: 
\begin{equation}
 \lambda = \frac{p}{|m|}(\frac{1}{\beta}-\cos\theta).
\label{eq:S_SPR}
\end{equation}

On the other hand, when the materials properties, $e.g.$, refractive index $n$ of a uniform dielectric slab, vary periodically with respect to time instead of space, a TG is created in \textcolor{blue}{Fig. \ref{fig1}(b)}. According to Bloch-Floquet theorem, the electric profiles of TG can be expanded by Fourier series: $E(t) = \sum_{m}e_{m}e^{j(\omega-m\Omega)t}$, where $\Omega=2\pi f_m$ with $f_m=1/T$ the modulation frequency and $T$ the modulation period, and $m=0,\pm1,\pm2,\cdots$ is the harmonic order, immediately suggesting that the temporal harmonic modulation will alter the operation frequency \cite{zurita2009reflection}.

When a free electron hovers above TG (grating/free-space interface at $z=0$) with a separation distance $d$, the excitation of the free electron is given by $J(x,z,t)=\hat{x}qv_{e}\delta(z-d)\delta(x-v_{e}t)$, and the frequency domain expression is obtained through Fourier transform $J(x,z,\omega)=\hat{x}q\delta(z-d) e^{j\omega t}e^{-jk_{xe}x}$, where $k_{xe} =\omega/v_{e}$. Hence, the evanescent wave induced by the free electron source can be written as:
\begin{equation}
H_{y}^{i} =\hat{y}A_{0}e^{j\omega t}e^{-jk_{xe}x-jk_{z0}(z-d)},
\label{eq:eigen}
\end{equation}
\begin{equation}
E_{x}^{i} =-\hat{x}A_{0}\frac{k_{xe}x}{\omega\varepsilon_{0}}e^{j\omega t}e^{-jk_{xe}x-jk_{z0}(z-d)}.
\label{eq:eigen}
\end{equation}
where $A_0$ is the electron amplitude. The reflected wave in the free space ($z>0$) is:
\begin{equation}
H_{y}^{r} =\hat{y}\sum_{m}R_{m}e^{j\omega t}e^{-jk_{xe}x-jk_{xm}z},
\label{eq:eigen}
\end{equation}
\begin{equation}
E_{x}^{r} =\hat{x}\sum_{m}R_{m}\frac{k_{xe}}{(\omega-m\Omega)\varepsilon_{0}}e^{j(\omega-m\Omega)t}e^{-jk_{xe}x-jk_{zm}z},
\label{eq:eigen}
\end{equation}
where  $k_{xe}^2+k_{zm}^2=k_{m}^2=[(\omega-m\Omega)/c]^2$. 
Eqs. (4)-(5) indicate that the frequency harmonic wave is induced by TG. Thus the electron dispersion line (red solid line) shifts either upward (for $m < 0$) or downward (for $m > 0$) along the $f$-axis in \textcolor{blue}{Fig. \ref{fig1}(d)}. Likewise, when the electron dispersion line moves above the light cone (red dashed line), the evanescent wave is converted into the plane wave in the far-field and gives rise to the $t$-SPR. 
The corresponding dispersion curve can be expressed as follows (see \textcolor{blue}{SI-1} \cite{supple} for derivation):
\begin{equation}
f =\frac{|m|}{T}\frac{1}{1-\beta \cos\theta}.
\label{eq6}
\end{equation}
This equation reveals that the frequency of $t$-SPR can be  customized by the modulation period $T$ and spans a frequency band for the $m^{\textrm{th}}$ frequency harmonic wave:
\begin{equation}
\frac{|m|}{1+\beta}\frac{1}{T}<f<\frac{|m|}{1-\beta}\frac{1}{T}.
\label{eq7}
\end{equation}

\begin{figure}[h]
\centering
\includegraphics[scale=1]{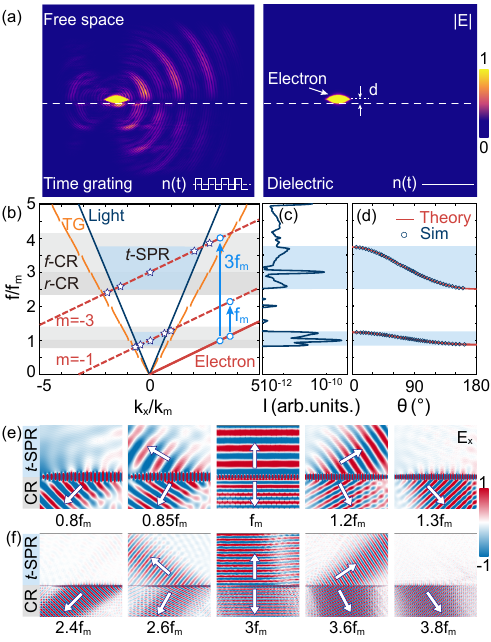}
\caption{\label{fig2}
Mechanism investigations on the simultaneous excitation of $t$-SPR and CR in an exemplified system with the normalized electron velocity $\beta = 0.2$ and square temporal modulation with $f_m=$ 20 GHz. 
(a) Field intensity profiles in time domain when a free electron flies over a dielectric slab with or without a temporal modulation. 
Left: $t$-SPR (and CR) is induced in the free space (and dielectric slab) with the temporal modulation. 
Right: no radiation is observed upon removal of temporal modulation. 
(b) The dispersion diagram of the exemplified system: light line (blue), energy band within TG (orange), electron (solid red), and electron harmonic wave (dashed red). 
The radiation frequency range can be determined by the intersection points between the electron harmonic wave and the light line for $t$-SPR (blue region); 
or between the electron harmonic wave and TG line for forward-CR ($f$-CR) and reverse-CR ($r$-CR) (grey regions). 
(c) The $t$-SPR radiation spectrum probed in the free space. (d) The relation between radiation angle ($\theta$) and frequency for $t$-SPR: 
theory (solid line) and simulations (symbols). 
(e)-(f) Simulated electric-field profiles at selected frequencies marked by stars in (b) for 1$^{\textrm{st}}$ ($m=-1$) and 3$^{\textrm{rd}}$ ($m=-3$) harmonic orders, where the arrows indicate the directions of $t$-SPR and CR in different scenarios.}
\end{figure}

To verify this dispersion relation, we run two dimensional time-domain numerical simulations and perform Fourier transformation (see \textcolor{blue}{SI-2} \cite{supple}) to vividly unveil the characteristics of $t$-SPR in \textcolor{blue}{Fig. \ref{fig2}}. 
We set our refractive index modulation function as $n(t) = A\cdot W(t) + n_a$, 
where $A$ denotes the amplitude, 
$W(t)$ represent the shape of the modulation function with $f_m=1/T$ the modulation frequency, 
and $n_a$ is the central refractive index.
We choose the following parameters throughout this study unless otherwise mentioned: $A = 0.2$, $n_a=1.5$, $f_m=20$ GHz, $W$ is in a square waveform; normalized electron velocity $\beta = 0.2$, and electron-grating separation $d = 0.5$ mm. 

As shown in \textcolor{blue}{Fig. \ref{fig2}(a)}, when the electron flies over the TG, it generates radiations both within free space and within the TG. Upon removing the temporal modulation, both radiations become imperceptible, indicating the crucial role played by temporal modulation.
To gain insight into the underlying mechanisms, we plot the dispersion diagram in \textcolor{blue}{Fig. \ref{fig2}(b)}. 
Here, the dispersion curve of TG (orange line) is derived by matching the temporal boundaries, which exhibits momentum gaps occurring at integer multiple frequencies of $0.5f_m$ (see \textcolor{blue}{SI-3} \cite{supple}). Within the momentum gap, the wave experiences exponential growth over time due to the extra energy injected into the system through modulation \cite{lustig2018topological,lyubarov2022amplified,dikopoltsev2022light}. 
The asymptotic light line in TG can also be expressed as $\omega=ck_{x}/n_{\textrm{eff}}$, with $n_{\textrm{eff}}$ denoting the effective refractive index of TG  \cite{pacheco2020effective}. 
Meanwhile, the frequency harmonic dispersion curve of the free electron shifts either upward or downward with distance $mf_m$ along the $f$-axis.
For example, the case of $m =-1$ is considered (red dashed); this $- 1^{\textrm{st}}$ harmonic wave transits above the free-space light cone, generating an overlapping area depicted as the blue region. 
It indicates the conversion of the evanescent wave of the free electron into a propagating wave in the far field, thereby giving rise to $t$-SPR. 
Simultaneously, this $- 1^{\textrm{st}}$ harmonic wave reaches above TG light cone, leading to an overlapping area depicted as the grey region, indicating the conversion into a propagating wave into TG.

\textcolor{blue}{Fig. \ref{fig2}(c)} presents the simulated radiation spectrum. As the TG shifts the electron harmonic line upward along the $f$-axis, the radiation characteristics at frequency $f$ rely on the evanescent wave at the frequency $f-f_{m}$. Serving as the intrinsic evanescent wave of the free electron wave, the decay distance increases as the frequency decreases while maintaining a constant velocity according to Bohr cutoff distance, $d_{\textrm{Bohr}}=\beta/(2\pi f \sqrt{1-\beta^2})$, the length scale over which the moving electron’s evanescent field decays exponentially with transverse distance from its axis of motion \cite{de2010optical, lu2023smith}. The electric intensity interacting with the TG becomes stronger as the operational frequency approaches zero. Consequently, the radiation spectrum reaches its peak at $f_m$ and subsequently diminishes as $f$ deviates from $f_m$.

In addition, Eq. (\ref{eq6}) predicts a relationship between the radiation angle ($\theta$) and frequency ($f$), which is shown in \textcolor{blue}{Fig. \ref{fig2}(d)} (line).
As suggested, $t$-SPR exhibits a beam-scanning characteristic within the operation bandwidth. 
The radiation direction switches from backward ($180^{\circ}$-$90^{\circ}$) to forward ($90^{\circ}$-$0^{\circ}$) as $f$ increases and a normal emission is produced at $f=f_m$. 
This relationship has been validated by our simulations in \textcolor{blue}{Fig. \ref{fig2}(d)} (symbols) and vividly illustrated by the electric-field profiles at selected $f$ in \textcolor{blue}{Fig. \ref{fig2}(e)}: 0.85$f_m$ (backward), $f_m$ (normal), 1.2$f_m$ (forward).
Moreover, $t$-SPR ceases to appear outside the predicted $t$-SPR range ($0.83f_m<f<1.25f_m$) according to Eq. (\ref{eq7}).

Interestingly, along with $t$-SPR excitation, there exist radiation within TG, which we account for CR due to the phase velocity of $-1^{\textrm{st}}$ harmonic wave exceeding that of light within TG as clearly evidenced in \textcolor{blue}{Fig. \ref{fig2}(b)}. 
The CR frequency range can be divided into two regions by the line $f=f_m$: reverse-CR ($r$-CR) and forward-CR ($f$-CR), which manifest in the left $(k_{x}<0)$ and right $(k_{x}>0)$ half-spaces, respectively, signifying the energy flow propagates backwardly or forwardly with respect to the electron motion. As depicted in \textcolor{blue}{Fig. \ref{fig2}(e)}, $r$-CR is evident at $0.8f_m$ and $0.85f_m$, while $f$-CR is observed at $1.2f_m$ and $1.3f_m$.
In particular, $r$-CR serves to separate electrons from the wake fields they generate, offering a wide range of applications in the high-energy particle field \cite{xi2009experimental,duan2017observation}. Traditionally, $r$-CR can be produced within double-negative metamaterials \cite{chen2011flipping}; here, the TG introduces a novel approach to its generation. 
It's important to note that the electron energy in our setup ($\beta=$ 0.2) is considerably lower than the minimum CR threshold $(\beta_{\textrm{CR}} = 1/1.7\approx0.59)$. 
Periodic temporal modulation provides the necessary energy boost for electrons to surpass the threshold required to induce CR \cite{dikopoltsev2022light}. It represents an exceptional solution to a long-standing challenge in eliminating the CR threshold besides the hyperbolic metamaterials \cite{liu2017integrated}. 
When electron energy is sufficiently high and surpasses the minimum energy threshold within TG ($e.g.,$ $\beta=0.6$), CR can be directly induced, both fundamental wave $(m=0)$ and higher-order harmonics $(|m|>0)$ come into play (see \textcolor{blue}{SI-4 \cite{supple}}).

Similar to conventional $s$-SPR, $t$-SPR also exhibits characteristics of high-order harmonic radiation. For instance, the $3^{\textrm{rd}}$ harmonic radiation around $f=3f_m$ is illustrated in \textcolor{blue}{Fig. \ref{fig2}(b)}. The radiation angle, as observed by the numerical simulations, closely aligns with theoretical expectations. However, in comparison to the $1^{\textrm{st}}$ order radiation, the intensity is noticeably diminished due to the declining harmonic intensity with increasing orders \cite{gaxiola2021temporal,gaxiola2023growing}. Nevertheless, it provides a potential pathway towards achieving frequency up-conversion and may find applications in the realm of extreme frequency lasing, especially in the ultraviolet range \cite{ye2019deep}. It's worth noting that the radiation spectrum near $2f_{m}$ is much weaker in this case, primarily because of the weak $2^{\textrm{nd}}$ harmonic amplitude \cite{gaxiola2021temporal}.

\begin{figure}[h]
\centering
\includegraphics[scale=1]{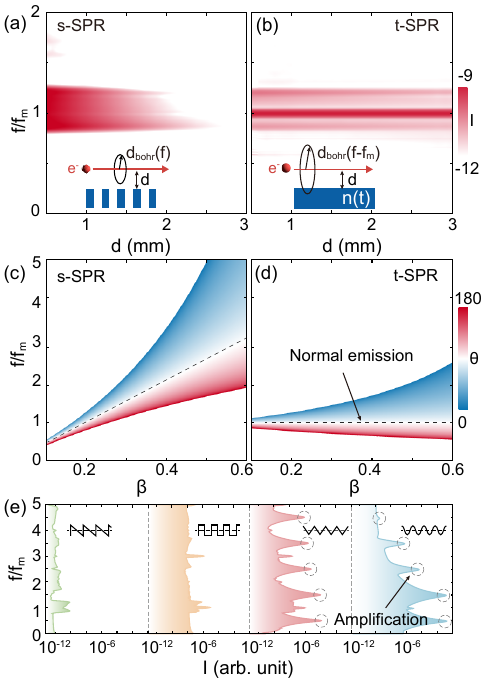}
\caption{\label{fig3}
Quantitative analysis for 1$^{\textrm{st}}$ order $s$-SPR and $t$-SPR.
The dependence of emission intensity spectra on electron-grating separation $d$ for (a) $s$-SPR or (b) $t$-SPR. 
Tuning radiation direction by varying electron velocity $\beta$ for (c) $s$-SPR or (d) $t$-SPR.
(e) Active control on $t$-SPR by changing the shape of temporal modulation function $W(t)$.
}
\end{figure}

In both $s$-SPR and $t$-SPR, the conversion of the electron's evanescent wave into free-space radiation is facilitated through the compensation of momentum or energy from the spatial or time grating, respectively. Their characteristics exhibit several distinctions, and a direct comparison is detailed in \textcolor{blue}{Fig. \ref{fig3}}. To ensure that $s$-SPR and $t$-SPR encompass the same radiation range, it is crucial to maintain a specific relationship between the spatial and temporal periods, denoted as $p = T\beta c$, while considering the same working harmonic order (see \textcolor{blue}{SI-5} \cite{supple}). It's worth noting that the evanescent field of a moving electron experiences exponential decay with increasing transverse distance, $i.e.$, $d_{\textrm{Bohr}}$, and the proximity of the electron trajectory to the grating surface $d$ plays a pivotal role \cite{lu2023smith}. In general, closer distances result in stronger interactions. However, this distance is constrained by experimental resolution to prevent direct electron impact on the structure for conventional $s$-SPR. 
In contrast, $t$-SPR demonstrates robustness to electron-grating separation $d$. 
With increased $d$, the intensity of $s$-SPR  exponentially decreases (\textcolor{blue}{Fig. \ref{fig3}(a)}), while that of $t$-SPR remains relatively constant (\textcolor{blue}{Fig. \ref{fig3}(b)}), which highlights their distinct excitation mechanisms. 
The intensity at the same radiation frequency $f$ is determined by the amplitude of frequency harmonic wave at $f-f_{m}$ for $t$-SPR but $f$ for $s$-SPR. Due to the much longer Bohr cutoff distance $d_{\textrm{Bohr}}(f-f_{m})$ compared to $d_{\textrm{Bohr}}(f)=0.48$ mm, $t$-SPR becomes relatively insensitive to electron-grating separation in the studied range of 0.5 -- 3 mm. This alleviates the constraints of electron operation within the near-field \cite{kong1975theory}. 
Additionally, we compare the radiation angle distributions for $s$-SPR and $t$-SPR in \textcolor{blue}{Fig. \ref{fig3}(c)-(d)} by tuning the normalized electron velocity $\beta$ and reveal that normal emission ($\theta=90^{\circ}$) consistently occurs at $f_m$ with varying $\beta$ for $t$-SPR.
This consistency is attributed to the fact that the radiation frequency $f_m$ remains steadfast along the $f$-axis, regardless of variations in $\beta$, and is accompanied by a corresponding wavenumber of $k_x = 0$. This feature highlights a potential application in frequency-locked light sources \cite{korbly2005observation,zhu2019free}.

Of particular significance, $t$-SPR offers several noteworthy advantages concerning energy and reconfigurability. \textcolor{blue}{First}, in $s$-SPR, a spatial grating functions as a passive system, adhering to the principles of energy conservation. It converts the evanescent waves carried by electrons into spatially propagating waves, with the total system energy being derived exclusively from the electrons. Conversely, $t$-SPR in TG operates as an active system, departing from strict adherence to energy conservation principles within the system. It has the capacity to introduce external energy through temporal modulation and transfer this energy to the electron-photon interaction, thus sidestepping conventional energy conservation constraints. Consequently, $t$-SPR introduces an additional energy channel with the potential to significantly enhance the radiation intensity. 
For example, when our external modulations are appropriately configured (the triangular or cosine modulation function as shown in \textcolor{blue}{Fig. \ref{fig3}(e)}), the amplification effect becomes evident at frequency $(2N+1)f_m/2$ where $N$ is an integer, consistent with findings from recent CR in PTC literature \cite{dikopoltsev2022light}. 
The leaking amplified wave contributes to the enhancement of $t$-SPR intensity. 
Functioning as an active amplification mechanism, $t$-SPR shows potential as a platform for on-chip electron accelerators with the precise matching of the phase conditions between the electron and the $t$-SPR wave \cite{zhang2023coherent,chlouba2023coherent,zhang2022coherent}.
\textcolor{blue}{Second}, $t$-SPR provides a reconfigurable platform. In $s$-SPR, the emission characteristics are contingent upon the electron energy with the fabricated grating. In contrast, the TG allows for adaptability through the adjustment of the modulation period $T$, amplitude $A$ (see \textcolor{blue}{SI-6} \cite{supple}), and modulation functions $W(t)$ \cite{sharabi2021disordered}. 
This attribute holds significant promises for the development of advanced active free-electron light sources and versatile system configurations.

In conclusion, we present a theoretical prediction of $t$-SPR from free electron traveling atop a simple TG. Looking forward, we aim to stimulate experimental verification with the key task of achieving high-speed modulation for the TG. 
Potential platforms for experimentation include graphene and epsilon-near-zero films, such as indium tin oxide \cite{phare2015graphene,sounas2017non}.
The consideration of TG with finite thickness is also paramount, and comprehensive information regarding this aspect can be found in \textcolor{blue}{SI-7} \cite{supple}. Analogous to $s$-SPR originating from a transmission grating, $t$-SPR can be detected in both the top and bottom half-spaces, following our derived radiation directions based on Eq. (\ref{eq6}).

\begin{acknowledgments}
This work was supported by the Singapore University of Technology and Design for the Start-Up Research Grant SRG SMT 2021 169 and Kickstarter Initiative (SKI)
SKI 2021-02-14, and National Research Foundation Singapore via Grant No. NRF2021-QEP2-02-P03, NRF2021-QEP2-03-P09, and NRF-CRP26-2021-0004.
\end{acknowledgments}

\bibliography{biblio}

\providecommand{\noopsort}[1]{}\providecommand{\singleletter}[1]{#1}%
\begin{thebibliography}{51}%
\makeatletter
\providecommand \@ifxundefined [1]{%
 \@ifx{#1\undefined}
}%
\providecommand \@ifnum [1]{%
 \ifnum #1\expandafter \@firstoftwo
 \else \expandafter \@secondoftwo
 \fi
}%
\providecommand \@ifx [1]{%
 \ifx #1\expandafter \@firstoftwo
 \else \expandafter \@secondoftwo
 \fi
}%
\providecommand \natexlab [1]{#1}%
\providecommand \enquote  [1]{``#1''}%
\providecommand \bibnamefont  [1]{#1}%
\providecommand \bibfnamefont [1]{#1}%
\providecommand \citenamefont [1]{#1}%
\providecommand \href@noop [0]{\@secondoftwo}%
\providecommand \href [0]{\begingroup \@sanitize@url \@href}%
\providecommand \@href[1]{\@@startlink{#1}\@@href}%
\providecommand \@@href[1]{\endgroup#1\@@endlink}%
\providecommand \@sanitize@url [0]{\catcode `\\12\catcode `\$12\catcode `\&12\catcode `\#12\catcode `\^12\catcode `\_12\catcode `\%12\relax}%
\providecommand \@@startlink[1]{}%
\providecommand \@@endlink[0]{}%
\providecommand \url  [0]{\begingroup\@sanitize@url \@url }%
\providecommand \@url [1]{\endgroup\@href {#1}{\urlprefix }}%
\providecommand \urlprefix  [0]{URL }%
\providecommand \Eprint [0]{\href }%
\providecommand \doibase [0]{http://dx.doi.org/}%
\providecommand \selectlanguage [0]{\@gobble}%
\providecommand \bibinfo  [0]{\@secondoftwo}%
\providecommand \bibfield  [0]{\@secondoftwo}%
\providecommand \translation [1]{[#1]}%
\providecommand \BibitemOpen [0]{}%
\providecommand \bibitemStop [0]{}%
\providecommand \bibitemNoStop [0]{.\EOS\space}%
\providecommand \EOS [0]{\spacefactor3000\relax}%
\providecommand \BibitemShut  [1]{\csname bibitem#1\endcsname}%
\let\auto@bib@innerbib\@empty
\bibitem [{\citenamefont {Smith}\ and\ \citenamefont {Purcell}(1953)}]{smith1953visible}%
  \BibitemOpen
  \bibfield  {author} {\bibinfo {author} {\bibfnamefont {S.~J.}\ \bibnamefont {Smith}}\ and\ \bibinfo {author} {\bibfnamefont {E.}~\bibnamefont {Purcell}},\ }\href@noop {} {\bibfield  {journal} {\bibinfo  {journal} {Physical Review}\ }\textbf {\bibinfo {volume} {92}},\ \bibinfo {pages} {1069} (\bibinfo {year} {1953})}\BibitemShut {NoStop}%
\bibitem [{\citenamefont {Leavitt}\ \emph {et~al.}(1979)\citenamefont {Leavitt}, \citenamefont {Wortman},\ and\ \citenamefont {Morrison}}]{leavitt1979orotron}%
  \BibitemOpen
  \bibfield  {author} {\bibinfo {author} {\bibfnamefont {R.~P.}\ \bibnamefont {Leavitt}}, \bibinfo {author} {\bibfnamefont {D.~E.}\ \bibnamefont {Wortman}}, \ and\ \bibinfo {author} {\bibfnamefont {C.~A.}\ \bibnamefont {Morrison}},\ }\href@noop {} {\bibfield  {journal} {\bibinfo  {journal} {Applied Physics Letters}\ }\textbf {\bibinfo {volume} {35}},\ \bibinfo {pages} {363} (\bibinfo {year} {1979})}\BibitemShut {NoStop}%
\bibitem [{\citenamefont {Nanni}\ \emph {et~al.}(2015)\citenamefont {Nanni}, \citenamefont {Huang}, \citenamefont {Hong}, \citenamefont {Ravi}, \citenamefont {Fallahi}, \citenamefont {Moriena}, \citenamefont {Dwayne~Miller},\ and\ \citenamefont {K{\"a}rtner}}]{nanni2015terahertz}%
  \BibitemOpen
  \bibfield  {author} {\bibinfo {author} {\bibfnamefont {E.~A.}\ \bibnamefont {Nanni}}, \bibinfo {author} {\bibfnamefont {W.~R.}\ \bibnamefont {Huang}}, \bibinfo {author} {\bibfnamefont {K.-H.}\ \bibnamefont {Hong}}, \bibinfo {author} {\bibfnamefont {K.}~\bibnamefont {Ravi}}, \bibinfo {author} {\bibfnamefont {A.}~\bibnamefont {Fallahi}}, \bibinfo {author} {\bibfnamefont {G.}~\bibnamefont {Moriena}}, \bibinfo {author} {\bibfnamefont {R.}~\bibnamefont {Dwayne~Miller}}, \ and\ \bibinfo {author} {\bibfnamefont {F.~X.}\ \bibnamefont {K{\"a}rtner}},\ }\href@noop {} {\bibfield  {journal} {\bibinfo  {journal} {Nature Communications}\ }\textbf {\bibinfo {volume} {6}},\ \bibinfo {pages} {8486} (\bibinfo {year} {2015})}\BibitemShut {NoStop}%
\bibitem [{\citenamefont {Urata}\ \emph {et~al.}(1998)\citenamefont {Urata}, \citenamefont {Goldstein}, \citenamefont {Kimmitt}, \citenamefont {Naumov}, \citenamefont {Platt},\ and\ \citenamefont {Walsh}}]{urata1998superradiant}%
  \BibitemOpen
  \bibfield  {author} {\bibinfo {author} {\bibfnamefont {J.}~\bibnamefont {Urata}}, \bibinfo {author} {\bibfnamefont {M.}~\bibnamefont {Goldstein}}, \bibinfo {author} {\bibfnamefont {M.}~\bibnamefont {Kimmitt}}, \bibinfo {author} {\bibfnamefont {A.}~\bibnamefont {Naumov}}, \bibinfo {author} {\bibfnamefont {C.}~\bibnamefont {Platt}}, \ and\ \bibinfo {author} {\bibfnamefont {J.}~\bibnamefont {Walsh}},\ }\href@noop {} {\bibfield  {journal} {\bibinfo  {journal} {Physical Review Letters}\ }\textbf {\bibinfo {volume} {80}},\ \bibinfo {pages} {516} (\bibinfo {year} {1998})}\BibitemShut {NoStop}%
\bibitem [{\citenamefont {Su}\ \emph {et~al.}(2019{\natexlab{a}})\citenamefont {Su}, \citenamefont {Xiong}, \citenamefont {Xu}, \citenamefont {Cai}, \citenamefont {Yin}, \citenamefont {Peng},\ and\ \citenamefont {Liu}}]{su2019manipulating}%
  \BibitemOpen
  \bibfield  {author} {\bibinfo {author} {\bibfnamefont {Z.}~\bibnamefont {Su}}, \bibinfo {author} {\bibfnamefont {B.}~\bibnamefont {Xiong}}, \bibinfo {author} {\bibfnamefont {Y.}~\bibnamefont {Xu}}, \bibinfo {author} {\bibfnamefont {Z.}~\bibnamefont {Cai}}, \bibinfo {author} {\bibfnamefont {J.}~\bibnamefont {Yin}}, \bibinfo {author} {\bibfnamefont {R.}~\bibnamefont {Peng}}, \ and\ \bibinfo {author} {\bibfnamefont {Y.}~\bibnamefont {Liu}},\ }\href@noop {} {\bibfield  {journal} {\bibinfo  {journal} {Advanced Optical Materials}\ }\textbf {\bibinfo {volume} {7}},\ \bibinfo {pages} {1801666} (\bibinfo {year} {2019}{\natexlab{a}})}\BibitemShut {NoStop}%
\bibitem [{\citenamefont {Roques-Carmes}\ \emph {et~al.}(2023)\citenamefont {Roques-Carmes}, \citenamefont {Kooi}, \citenamefont {Yang}, \citenamefont {Rivera}, \citenamefont {Keathley}, \citenamefont {Joannopoulos}, \citenamefont {Johnson}, \citenamefont {Kaminer}, \citenamefont {Berggren},\ and\ \citenamefont {Solja{\v{c}}i{\'c}}}]{roques2023free}%
  \BibitemOpen
  \bibfield  {author} {\bibinfo {author} {\bibfnamefont {C.}~\bibnamefont {Roques-Carmes}}, \bibinfo {author} {\bibfnamefont {S.~E.}\ \bibnamefont {Kooi}}, \bibinfo {author} {\bibfnamefont {Y.}~\bibnamefont {Yang}}, \bibinfo {author} {\bibfnamefont {N.}~\bibnamefont {Rivera}}, \bibinfo {author} {\bibfnamefont {P.~D.}\ \bibnamefont {Keathley}}, \bibinfo {author} {\bibfnamefont {J.~D.}\ \bibnamefont {Joannopoulos}}, \bibinfo {author} {\bibfnamefont {S.~G.}\ \bibnamefont {Johnson}}, \bibinfo {author} {\bibfnamefont {I.}~\bibnamefont {Kaminer}}, \bibinfo {author} {\bibfnamefont {K.~K.}\ \bibnamefont {Berggren}}, \ and\ \bibinfo {author} {\bibfnamefont {M.}~\bibnamefont {Solja{\v{c}}i{\'c}}},\ }\href@noop {} {\bibfield  {journal} {\bibinfo  {journal} {Applied Physics Reviews}\ }\textbf {\bibinfo {volume} {10}} (\bibinfo {year} {2023})}\BibitemShut {NoStop}%
\bibitem [{\citenamefont {Korbly}\ \emph {et~al.}(2005)\citenamefont {Korbly}, \citenamefont {Kesar}, \citenamefont {Sirigiri},\ and\ \citenamefont {Temkin}}]{korbly2005observation}%
  \BibitemOpen
  \bibfield  {author} {\bibinfo {author} {\bibfnamefont {S.}~\bibnamefont {Korbly}}, \bibinfo {author} {\bibfnamefont {A.}~\bibnamefont {Kesar}}, \bibinfo {author} {\bibfnamefont {J.}~\bibnamefont {Sirigiri}}, \ and\ \bibinfo {author} {\bibfnamefont {R.}~\bibnamefont {Temkin}},\ }\href@noop {} {\bibfield  {journal} {\bibinfo  {journal} {Physical Review Letters}\ }\textbf {\bibinfo {volume} {94}},\ \bibinfo {pages} {054803} (\bibinfo {year} {2005})}\BibitemShut {NoStop}%
\bibitem [{\citenamefont {Wang}\ \emph {et~al.}(2016)\citenamefont {Wang}, \citenamefont {Yao}, \citenamefont {Chen}, \citenamefont {Chen},\ and\ \citenamefont {Liu}}]{wang2016manipulating}%
  \BibitemOpen
  \bibfield  {author} {\bibinfo {author} {\bibfnamefont {Z.}~\bibnamefont {Wang}}, \bibinfo {author} {\bibfnamefont {K.}~\bibnamefont {Yao}}, \bibinfo {author} {\bibfnamefont {M.}~\bibnamefont {Chen}}, \bibinfo {author} {\bibfnamefont {H.}~\bibnamefont {Chen}}, \ and\ \bibinfo {author} {\bibfnamefont {Y.}~\bibnamefont {Liu}},\ }\href@noop {} {\bibfield  {journal} {\bibinfo  {journal} {Physical Review Letters}\ }\textbf {\bibinfo {volume} {117}},\ \bibinfo {pages} {157401} (\bibinfo {year} {2016})}\BibitemShut {NoStop}%
\bibitem [{\citenamefont {Su}\ \emph {et~al.}(2019{\natexlab{b}})\citenamefont {Su}, \citenamefont {Cheng}, \citenamefont {Li},\ and\ \citenamefont {Liu}}]{su2019complete}%
  \BibitemOpen
  \bibfield  {author} {\bibinfo {author} {\bibfnamefont {Z.}~\bibnamefont {Su}}, \bibinfo {author} {\bibfnamefont {F.}~\bibnamefont {Cheng}}, \bibinfo {author} {\bibfnamefont {L.}~\bibnamefont {Li}}, \ and\ \bibinfo {author} {\bibfnamefont {Y.}~\bibnamefont {Liu}},\ }\href@noop {} {\bibfield  {journal} {\bibinfo  {journal} {ACS Photonics}\ }\textbf {\bibinfo {volume} {6}},\ \bibinfo {pages} {1947} (\bibinfo {year} {2019}{\natexlab{b}})}\BibitemShut {NoStop}%
\bibitem [{\citenamefont {Kaminer}\ \emph {et~al.}(2017)\citenamefont {Kaminer}, \citenamefont {Kooi}, \citenamefont {Shiloh}, \citenamefont {Zhen}, \citenamefont {Shen}, \citenamefont {L{\'o}pez}, \citenamefont {Remez}, \citenamefont {Skirlo}, \citenamefont {Yang}, \citenamefont {Joannopoulos} \emph {et~al.}}]{kaminer2017spectrally}%
  \BibitemOpen
  \bibfield  {author} {\bibinfo {author} {\bibfnamefont {I.}~\bibnamefont {Kaminer}}, \bibinfo {author} {\bibfnamefont {S.}~\bibnamefont {Kooi}}, \bibinfo {author} {\bibfnamefont {R.}~\bibnamefont {Shiloh}}, \bibinfo {author} {\bibfnamefont {B.}~\bibnamefont {Zhen}}, \bibinfo {author} {\bibfnamefont {Y.}~\bibnamefont {Shen}}, \bibinfo {author} {\bibfnamefont {J.}~\bibnamefont {L{\'o}pez}}, \bibinfo {author} {\bibfnamefont {R.}~\bibnamefont {Remez}}, \bibinfo {author} {\bibfnamefont {S.}~\bibnamefont {Skirlo}}, \bibinfo {author} {\bibfnamefont {Y.}~\bibnamefont {Yang}}, \bibinfo {author} {\bibfnamefont {J.}~\bibnamefont {Joannopoulos}},  \emph {et~al.},\ }\href@noop {} {\bibfield  {journal} {\bibinfo  {journal} {Physical Review X}\ }\textbf {\bibinfo {volume} {7}},\ \bibinfo {pages} {011003} (\bibinfo {year} {2017})}\BibitemShut {NoStop}%
\bibitem [{\citenamefont {Jing}\ \emph {et~al.}(2021)\citenamefont {Jing}, \citenamefont {Lin}, \citenamefont {Wang}, \citenamefont {Kaminer}, \citenamefont {Hu}, \citenamefont {Li}, \citenamefont {Liu}, \citenamefont {Chen}, \citenamefont {Zhang},\ and\ \citenamefont {Chen}}]{jing2021polarization}%
  \BibitemOpen
  \bibfield  {author} {\bibinfo {author} {\bibfnamefont {L.}~\bibnamefont {Jing}}, \bibinfo {author} {\bibfnamefont {X.}~\bibnamefont {Lin}}, \bibinfo {author} {\bibfnamefont {Z.}~\bibnamefont {Wang}}, \bibinfo {author} {\bibfnamefont {I.}~\bibnamefont {Kaminer}}, \bibinfo {author} {\bibfnamefont {H.}~\bibnamefont {Hu}}, \bibinfo {author} {\bibfnamefont {E.}~\bibnamefont {Li}}, \bibinfo {author} {\bibfnamefont {Y.}~\bibnamefont {Liu}}, \bibinfo {author} {\bibfnamefont {M.}~\bibnamefont {Chen}}, \bibinfo {author} {\bibfnamefont {B.}~\bibnamefont {Zhang}}, \ and\ \bibinfo {author} {\bibfnamefont {H.}~\bibnamefont {Chen}},\ }\href@noop {} {\bibfield  {journal} {\bibinfo  {journal} {Laser \& Photonics Reviews}\ }\textbf {\bibinfo {volume} {15}},\ \bibinfo {pages} {2000426} (\bibinfo {year} {2021})}\BibitemShut {NoStop}%
\bibitem [{\citenamefont {Jing}\ \emph {et~al.}(2019)\citenamefont {Jing}, \citenamefont {Wang}, \citenamefont {Lin}, \citenamefont {Zheng}, \citenamefont {Xu}, \citenamefont {Shen}, \citenamefont {Yang}, \citenamefont {Gao}, \citenamefont {Chen},\ and\ \citenamefont {Chen}}]{jing2019spiral}%
  \BibitemOpen
  \bibfield  {author} {\bibinfo {author} {\bibfnamefont {L.}~\bibnamefont {Jing}}, \bibinfo {author} {\bibfnamefont {Z.}~\bibnamefont {Wang}}, \bibinfo {author} {\bibfnamefont {X.}~\bibnamefont {Lin}}, \bibinfo {author} {\bibfnamefont {B.}~\bibnamefont {Zheng}}, \bibinfo {author} {\bibfnamefont {S.}~\bibnamefont {Xu}}, \bibinfo {author} {\bibfnamefont {L.}~\bibnamefont {Shen}}, \bibinfo {author} {\bibfnamefont {Y.}~\bibnamefont {Yang}}, \bibinfo {author} {\bibfnamefont {F.}~\bibnamefont {Gao}}, \bibinfo {author} {\bibfnamefont {M.}~\bibnamefont {Chen}}, \ and\ \bibinfo {author} {\bibfnamefont {H.}~\bibnamefont {Chen}},\ }\href@noop {} {\bibfield  {journal} {\bibinfo  {journal} {Research}\ } (\bibinfo {year} {2019})}\BibitemShut {NoStop}%
\bibitem [{\citenamefont {Zhang}\ \emph {et~al.}(2023{\natexlab{a}})\citenamefont {Zhang}, \citenamefont {Zhu}, \citenamefont {Du}, \citenamefont {Gao}, \citenamefont {Han},\ and\ \citenamefont {Liu}}]{zhang2023chiral}%
  \BibitemOpen
  \bibfield  {author} {\bibinfo {author} {\bibfnamefont {Z.-W.}\ \bibnamefont {Zhang}}, \bibinfo {author} {\bibfnamefont {J.-F.}\ \bibnamefont {Zhu}}, \bibinfo {author} {\bibfnamefont {C.-H.}\ \bibnamefont {Du}}, \bibinfo {author} {\bibfnamefont {F.}~\bibnamefont {Gao}}, \bibinfo {author} {\bibfnamefont {F.-Y.}\ \bibnamefont {Han}}, \ and\ \bibinfo {author} {\bibfnamefont {P.-K.}\ \bibnamefont {Liu}},\ }\href@noop {} {\bibfield  {journal} {\bibinfo  {journal} {Laser \& Photonics Reviews}\ }\textbf {\bibinfo {volume} {17}},\ \bibinfo {pages} {2200420} (\bibinfo {year} {2023}{\natexlab{a}})}\BibitemShut {NoStop}%
\bibitem [{\citenamefont {Yang}\ \emph {et~al.}(2018)\citenamefont {Yang}, \citenamefont {Massuda}, \citenamefont {Roques-Carmes}, \citenamefont {Kooi}, \citenamefont {Christensen}, \citenamefont {Johnson}, \citenamefont {Joannopoulos}, \citenamefont {Miller}, \citenamefont {Kaminer},\ and\ \citenamefont {Solja{\v{c}}i{\'c}}}]{yang2018maximal}%
  \BibitemOpen
  \bibfield  {author} {\bibinfo {author} {\bibfnamefont {Y.}~\bibnamefont {Yang}}, \bibinfo {author} {\bibfnamefont {A.}~\bibnamefont {Massuda}}, \bibinfo {author} {\bibfnamefont {C.}~\bibnamefont {Roques-Carmes}}, \bibinfo {author} {\bibfnamefont {S.~E.}\ \bibnamefont {Kooi}}, \bibinfo {author} {\bibfnamefont {T.}~\bibnamefont {Christensen}}, \bibinfo {author} {\bibfnamefont {S.~G.}\ \bibnamefont {Johnson}}, \bibinfo {author} {\bibfnamefont {J.~D.}\ \bibnamefont {Joannopoulos}}, \bibinfo {author} {\bibfnamefont {O.~D.}\ \bibnamefont {Miller}}, \bibinfo {author} {\bibfnamefont {I.}~\bibnamefont {Kaminer}}, \ and\ \bibinfo {author} {\bibfnamefont {M.}~\bibnamefont {Solja{\v{c}}i{\'c}}},\ }\href@noop {} {\bibfield  {journal} {\bibinfo  {journal} {Nature Physics}\ }\textbf {\bibinfo {volume} {14}},\ \bibinfo {pages} {894} (\bibinfo {year} {2018})}\BibitemShut {NoStop}%
\bibitem [{\citenamefont {Yang}\ \emph {et~al.}(2023)\citenamefont {Yang}, \citenamefont {Roques-Carmes}, \citenamefont {Kooi}, \citenamefont {Tang}, \citenamefont {Beroz}, \citenamefont {Mazur}, \citenamefont {Kaminer}, \citenamefont {Joannopoulos},\ and\ \citenamefont {Solja{\v{c}}i{\'c}}}]{yang2023photonic}%
  \BibitemOpen
  \bibfield  {author} {\bibinfo {author} {\bibfnamefont {Y.}~\bibnamefont {Yang}}, \bibinfo {author} {\bibfnamefont {C.}~\bibnamefont {Roques-Carmes}}, \bibinfo {author} {\bibfnamefont {S.~E.}\ \bibnamefont {Kooi}}, \bibinfo {author} {\bibfnamefont {H.}~\bibnamefont {Tang}}, \bibinfo {author} {\bibfnamefont {J.}~\bibnamefont {Beroz}}, \bibinfo {author} {\bibfnamefont {E.}~\bibnamefont {Mazur}}, \bibinfo {author} {\bibfnamefont {I.}~\bibnamefont {Kaminer}}, \bibinfo {author} {\bibfnamefont {J.~D.}\ \bibnamefont {Joannopoulos}}, \ and\ \bibinfo {author} {\bibfnamefont {M.}~\bibnamefont {Solja{\v{c}}i{\'c}}},\ }\href@noop {} {\bibfield  {journal} {\bibinfo  {journal} {Nature}\ }\textbf {\bibinfo {volume} {613}},\ \bibinfo {pages} {42} (\bibinfo {year} {2023})}\BibitemShut {NoStop}%
\bibitem [{\citenamefont {Engheta}(2023)}]{engheta2023four}%
  \BibitemOpen
  \bibfield  {author} {\bibinfo {author} {\bibfnamefont {N.}~\bibnamefont {Engheta}},\ }\href@noop {} {\bibfield  {journal} {\bibinfo  {journal} {Science}\ }\textbf {\bibinfo {volume} {379}},\ \bibinfo {pages} {1190} (\bibinfo {year} {2023})}\BibitemShut {NoStop}%
\bibitem [{\citenamefont {Fante}(1971)}]{fante1971transmission}%
  \BibitemOpen
  \bibfield  {author} {\bibinfo {author} {\bibfnamefont {R.}~\bibnamefont {Fante}},\ }\href@noop {} {\bibfield  {journal} {\bibinfo  {journal} {IEEE Transactions on Antennas and Propagation}\ }\textbf {\bibinfo {volume} {19}},\ \bibinfo {pages} {417} (\bibinfo {year} {1971})}\BibitemShut {NoStop}%
\bibitem [{\citenamefont {Moussa}\ \emph {et~al.}(2023)\citenamefont {Moussa}, \citenamefont {Xu}, \citenamefont {Yin}, \citenamefont {Galiffi}, \citenamefont {Ra’di},\ and\ \citenamefont {Al{\`u}}}]{moussa2023observation}%
  \BibitemOpen
  \bibfield  {author} {\bibinfo {author} {\bibfnamefont {H.}~\bibnamefont {Moussa}}, \bibinfo {author} {\bibfnamefont {G.}~\bibnamefont {Xu}}, \bibinfo {author} {\bibfnamefont {S.}~\bibnamefont {Yin}}, \bibinfo {author} {\bibfnamefont {E.}~\bibnamefont {Galiffi}}, \bibinfo {author} {\bibfnamefont {Y.}~\bibnamefont {Ra’di}}, \ and\ \bibinfo {author} {\bibfnamefont {A.}~\bibnamefont {Al{\`u}}},\ }\href@noop {} {\bibfield  {journal} {\bibinfo  {journal} {Nature Physics}\ }\textbf {\bibinfo {volume} {19}},\ \bibinfo {pages} {863} (\bibinfo {year} {2023})}\BibitemShut {NoStop}%
\bibitem [{\citenamefont {Bacot}\ \emph {et~al.}(2016)\citenamefont {Bacot}, \citenamefont {Labousse}, \citenamefont {Eddi}, \citenamefont {Fink},\ and\ \citenamefont {Fort}}]{bacot2016time}%
  \BibitemOpen
  \bibfield  {author} {\bibinfo {author} {\bibfnamefont {V.}~\bibnamefont {Bacot}}, \bibinfo {author} {\bibfnamefont {M.}~\bibnamefont {Labousse}}, \bibinfo {author} {\bibfnamefont {A.}~\bibnamefont {Eddi}}, \bibinfo {author} {\bibfnamefont {M.}~\bibnamefont {Fink}}, \ and\ \bibinfo {author} {\bibfnamefont {E.}~\bibnamefont {Fort}},\ }\href@noop {} {\bibfield  {journal} {\bibinfo  {journal} {Nature Physics}\ }\textbf {\bibinfo {volume} {12}},\ \bibinfo {pages} {972} (\bibinfo {year} {2016})}\BibitemShut {NoStop}%
\bibitem [{\citenamefont {Sounas}\ and\ \citenamefont {Al{\`u}}(2017)}]{sounas2017non}%
  \BibitemOpen
  \bibfield  {author} {\bibinfo {author} {\bibfnamefont {D.~L.}\ \bibnamefont {Sounas}}\ and\ \bibinfo {author} {\bibfnamefont {A.}~\bibnamefont {Al{\`u}}},\ }\href@noop {} {\bibfield  {journal} {\bibinfo  {journal} {Nature Photonics}\ }\textbf {\bibinfo {volume} {11}},\ \bibinfo {pages} {774} (\bibinfo {year} {2017})}\BibitemShut {NoStop}%
\bibitem [{\citenamefont {Guo}\ \emph {et~al.}(2019)\citenamefont {Guo}, \citenamefont {Ding}, \citenamefont {Duan},\ and\ \citenamefont {Ni}}]{guo2019nonreciprocal}%
  \BibitemOpen
  \bibfield  {author} {\bibinfo {author} {\bibfnamefont {X.}~\bibnamefont {Guo}}, \bibinfo {author} {\bibfnamefont {Y.}~\bibnamefont {Ding}}, \bibinfo {author} {\bibfnamefont {Y.}~\bibnamefont {Duan}}, \ and\ \bibinfo {author} {\bibfnamefont {X.}~\bibnamefont {Ni}},\ }\href@noop {} {\bibfield  {journal} {\bibinfo  {journal} {Light: Science \& Applications}\ }\textbf {\bibinfo {volume} {8}},\ \bibinfo {pages} {123} (\bibinfo {year} {2019})}\BibitemShut {NoStop}%
\bibitem [{\citenamefont {Pacheco-Pe{\~n}a}\ and\ \citenamefont {Engheta}(2020{\natexlab{a}})}]{pacheco2020temporal}%
  \BibitemOpen
  \bibfield  {author} {\bibinfo {author} {\bibfnamefont {V.}~\bibnamefont {Pacheco-Pe{\~n}a}}\ and\ \bibinfo {author} {\bibfnamefont {N.}~\bibnamefont {Engheta}},\ }\href@noop {} {\bibfield  {journal} {\bibinfo  {journal} {Light: Science \& Applications}\ }\textbf {\bibinfo {volume} {9}},\ \bibinfo {pages} {129} (\bibinfo {year} {2020}{\natexlab{a}})}\BibitemShut {NoStop}%
\bibitem [{\citenamefont {Pacheco-Pe{\~n}a}\ and\ \citenamefont {Engheta}(2020{\natexlab{b}})}]{pacheco2020antireflection}%
  \BibitemOpen
  \bibfield  {author} {\bibinfo {author} {\bibfnamefont {V.}~\bibnamefont {Pacheco-Pe{\~n}a}}\ and\ \bibinfo {author} {\bibfnamefont {N.}~\bibnamefont {Engheta}},\ }\href@noop {} {\bibfield  {journal} {\bibinfo  {journal} {Optica}\ }\textbf {\bibinfo {volume} {7}},\ \bibinfo {pages} {323} (\bibinfo {year} {2020}{\natexlab{b}})}\BibitemShut {NoStop}%
\bibitem [{\citenamefont {Tirole}\ \emph {et~al.}(2023)\citenamefont {Tirole}, \citenamefont {Vezzoli}, \citenamefont {Galiffi}, \citenamefont {Robertson}, \citenamefont {Maurice}, \citenamefont {Tilmann}, \citenamefont {Maier}, \citenamefont {Pendry},\ and\ \citenamefont {Sapienza}}]{tirole2023double}%
  \BibitemOpen
  \bibfield  {author} {\bibinfo {author} {\bibfnamefont {R.}~\bibnamefont {Tirole}}, \bibinfo {author} {\bibfnamefont {S.}~\bibnamefont {Vezzoli}}, \bibinfo {author} {\bibfnamefont {E.}~\bibnamefont {Galiffi}}, \bibinfo {author} {\bibfnamefont {I.}~\bibnamefont {Robertson}}, \bibinfo {author} {\bibfnamefont {D.}~\bibnamefont {Maurice}}, \bibinfo {author} {\bibfnamefont {B.}~\bibnamefont {Tilmann}}, \bibinfo {author} {\bibfnamefont {S.~A.}\ \bibnamefont {Maier}}, \bibinfo {author} {\bibfnamefont {J.~B.}\ \bibnamefont {Pendry}}, \ and\ \bibinfo {author} {\bibfnamefont {R.}~\bibnamefont {Sapienza}},\ }\href@noop {} {\bibfield  {journal} {\bibinfo  {journal} {Nature Physics}\ }\textbf {\bibinfo {volume} {19}},\ \bibinfo {pages} {999–1002} (\bibinfo {year} {2023})}\BibitemShut {NoStop}%
\bibitem [{\citenamefont {Galiffi}\ \emph {et~al.}(2023)\citenamefont {Galiffi}, \citenamefont {Xu}, \citenamefont {Yin}, \citenamefont {Moussa}, \citenamefont {Ra’di},\ and\ \citenamefont {Al{\`u}}}]{galiffi2023broadband}%
  \BibitemOpen
  \bibfield  {author} {\bibinfo {author} {\bibfnamefont {E.}~\bibnamefont {Galiffi}}, \bibinfo {author} {\bibfnamefont {G.}~\bibnamefont {Xu}}, \bibinfo {author} {\bibfnamefont {S.}~\bibnamefont {Yin}}, \bibinfo {author} {\bibfnamefont {H.}~\bibnamefont {Moussa}}, \bibinfo {author} {\bibfnamefont {Y.}~\bibnamefont {Ra’di}}, \ and\ \bibinfo {author} {\bibfnamefont {A.}~\bibnamefont {Al{\`u}}},\ }\href@noop {} {\bibfield  {journal} {\bibinfo  {journal} {Nature Physics}\ }\textbf {\bibinfo {volume} {19}},\ \bibinfo {pages} {1703–1708} (\bibinfo {year} {2023})}\BibitemShut {NoStop}%
\bibitem [{\citenamefont {Lee}\ \emph {et~al.}(2018)\citenamefont {Lee}, \citenamefont {Son}, \citenamefont {Park}, \citenamefont {Kang}, \citenamefont {Jeon}, \citenamefont {Rotermund},\ and\ \citenamefont {Min}}]{lee2018linear}%
  \BibitemOpen
  \bibfield  {author} {\bibinfo {author} {\bibfnamefont {K.}~\bibnamefont {Lee}}, \bibinfo {author} {\bibfnamefont {J.}~\bibnamefont {Son}}, \bibinfo {author} {\bibfnamefont {J.}~\bibnamefont {Park}}, \bibinfo {author} {\bibfnamefont {B.}~\bibnamefont {Kang}}, \bibinfo {author} {\bibfnamefont {W.}~\bibnamefont {Jeon}}, \bibinfo {author} {\bibfnamefont {F.}~\bibnamefont {Rotermund}}, \ and\ \bibinfo {author} {\bibfnamefont {B.}~\bibnamefont {Min}},\ }\href@noop {} {\bibfield  {journal} {\bibinfo  {journal} {Nature Photonics}\ }\textbf {\bibinfo {volume} {12}},\ \bibinfo {pages} {765} (\bibinfo {year} {2018})}\BibitemShut {NoStop}%
\bibitem [{\citenamefont {Oue}\ \emph {et~al.}(2022)\citenamefont {Oue}, \citenamefont {Ding},\ and\ \citenamefont {Pendry}}]{oue2022vcerenkov}%
  \BibitemOpen
  \bibfield  {author} {\bibinfo {author} {\bibfnamefont {D.}~\bibnamefont {Oue}}, \bibinfo {author} {\bibfnamefont {K.}~\bibnamefont {Ding}}, \ and\ \bibinfo {author} {\bibfnamefont {J.}~\bibnamefont {Pendry}},\ }\href@noop {} {\bibfield  {journal} {\bibinfo  {journal} {Physical Review Research}\ }\textbf {\bibinfo {volume} {4}},\ \bibinfo {pages} {013064} (\bibinfo {year} {2022})}\BibitemShut {NoStop}%
\bibitem [{\citenamefont {Lustig}\ \emph {et~al.}(2018)\citenamefont {Lustig}, \citenamefont {Sharabi},\ and\ \citenamefont {Segev}}]{lustig2018topological}%
  \BibitemOpen
  \bibfield  {author} {\bibinfo {author} {\bibfnamefont {E.}~\bibnamefont {Lustig}}, \bibinfo {author} {\bibfnamefont {Y.}~\bibnamefont {Sharabi}}, \ and\ \bibinfo {author} {\bibfnamefont {M.}~\bibnamefont {Segev}},\ }\href@noop {} {\bibfield  {journal} {\bibinfo  {journal} {Optica}\ }\textbf {\bibinfo {volume} {5}},\ \bibinfo {pages} {1390} (\bibinfo {year} {2018})}\BibitemShut {NoStop}%
\bibitem [{\citenamefont {Lyubarov}\ \emph {et~al.}(2022)\citenamefont {Lyubarov}, \citenamefont {Lumer}, \citenamefont {Dikopoltsev}, \citenamefont {Lustig}, \citenamefont {Sharabi},\ and\ \citenamefont {Segev}}]{lyubarov2022amplified}%
  \BibitemOpen
  \bibfield  {author} {\bibinfo {author} {\bibfnamefont {M.}~\bibnamefont {Lyubarov}}, \bibinfo {author} {\bibfnamefont {Y.}~\bibnamefont {Lumer}}, \bibinfo {author} {\bibfnamefont {A.}~\bibnamefont {Dikopoltsev}}, \bibinfo {author} {\bibfnamefont {E.}~\bibnamefont {Lustig}}, \bibinfo {author} {\bibfnamefont {Y.}~\bibnamefont {Sharabi}}, \ and\ \bibinfo {author} {\bibfnamefont {M.}~\bibnamefont {Segev}},\ }\href@noop {} {\bibfield  {journal} {\bibinfo  {journal} {Science}\ }\textbf {\bibinfo {volume} {377}},\ \bibinfo {pages} {425} (\bibinfo {year} {2022})}\BibitemShut {NoStop}%
\bibitem [{\citenamefont {Wang}\ \emph {et~al.}(2023)\citenamefont {Wang}, \citenamefont {Mirmoosa}, \citenamefont {Asadchy}, \citenamefont {Rockstuhl}, \citenamefont {Fan},\ and\ \citenamefont {Tretyakov}}]{wang2023metasurface}%
  \BibitemOpen
  \bibfield  {author} {\bibinfo {author} {\bibfnamefont {X.}~\bibnamefont {Wang}}, \bibinfo {author} {\bibfnamefont {M.~S.}\ \bibnamefont {Mirmoosa}}, \bibinfo {author} {\bibfnamefont {V.~S.}\ \bibnamefont {Asadchy}}, \bibinfo {author} {\bibfnamefont {C.}~\bibnamefont {Rockstuhl}}, \bibinfo {author} {\bibfnamefont {S.}~\bibnamefont {Fan}}, \ and\ \bibinfo {author} {\bibfnamefont {S.~A.}\ \bibnamefont {Tretyakov}},\ }\href@noop {} {\bibfield  {journal} {\bibinfo  {journal} {Science Advances}\ }\textbf {\bibinfo {volume} {9}},\ \bibinfo {pages} {eadg7541} (\bibinfo {year} {2023})}\BibitemShut {NoStop}%
\bibitem [{\citenamefont {Galiffi}\ \emph {et~al.}(2020)\citenamefont {Galiffi}, \citenamefont {Wang}, \citenamefont {Lim}, \citenamefont {Pendry}, \citenamefont {Al{\`u}},\ and\ \citenamefont {Huidobro}}]{galiffi2020wood}%
  \BibitemOpen
  \bibfield  {author} {\bibinfo {author} {\bibfnamefont {E.}~\bibnamefont {Galiffi}}, \bibinfo {author} {\bibfnamefont {Y.-T.}\ \bibnamefont {Wang}}, \bibinfo {author} {\bibfnamefont {Z.}~\bibnamefont {Lim}}, \bibinfo {author} {\bibfnamefont {J.~B.}\ \bibnamefont {Pendry}}, \bibinfo {author} {\bibfnamefont {A.}~\bibnamefont {Al{\`u}}}, \ and\ \bibinfo {author} {\bibfnamefont {P.~A.}\ \bibnamefont {Huidobro}},\ }\href@noop {} {\bibfield  {journal} {\bibinfo  {journal} {Physical Review Letters}\ }\textbf {\bibinfo {volume} {125}},\ \bibinfo {pages} {127403} (\bibinfo {year} {2020})}\BibitemShut {NoStop}%
\bibitem [{\citenamefont {Dikopoltsev}\ \emph {et~al.}(2022)\citenamefont {Dikopoltsev}, \citenamefont {Sharabi}, \citenamefont {Lyubarov}, \citenamefont {Lumer}, \citenamefont {Tsesses}, \citenamefont {Lustig}, \citenamefont {Kaminer},\ and\ \citenamefont {Segev}}]{dikopoltsev2022light}%
  \BibitemOpen
  \bibfield  {author} {\bibinfo {author} {\bibfnamefont {A.}~\bibnamefont {Dikopoltsev}}, \bibinfo {author} {\bibfnamefont {Y.}~\bibnamefont {Sharabi}}, \bibinfo {author} {\bibfnamefont {M.}~\bibnamefont {Lyubarov}}, \bibinfo {author} {\bibfnamefont {Y.}~\bibnamefont {Lumer}}, \bibinfo {author} {\bibfnamefont {S.}~\bibnamefont {Tsesses}}, \bibinfo {author} {\bibfnamefont {E.}~\bibnamefont {Lustig}}, \bibinfo {author} {\bibfnamefont {I.}~\bibnamefont {Kaminer}}, \ and\ \bibinfo {author} {\bibfnamefont {M.}~\bibnamefont {Segev}},\ }\href@noop {} {\bibfield  {journal} {\bibinfo  {journal} {Proceedings of the National Academy of Sciences}\ }\textbf {\bibinfo {volume} {119}},\ \bibinfo {pages} {e2119705119} (\bibinfo {year} {2022})}\BibitemShut {NoStop}%
\bibitem [{\citenamefont {Zurita-S{\'a}nchez}\ \emph {et~al.}(2009)\citenamefont {Zurita-S{\'a}nchez}, \citenamefont {Halevi},\ and\ \citenamefont {Cervantes-Gonz{\'a}lez}}]{zurita2009reflection}%
  \BibitemOpen
  \bibfield  {author} {\bibinfo {author} {\bibfnamefont {J.~R.}\ \bibnamefont {Zurita-S{\'a}nchez}}, \bibinfo {author} {\bibfnamefont {P.}~\bibnamefont {Halevi}}, \ and\ \bibinfo {author} {\bibfnamefont {J.~C.}\ \bibnamefont {Cervantes-Gonz{\'a}lez}},\ }\href@noop {} {\bibfield  {journal} {\bibinfo  {journal} {Physical Review A}\ }\textbf {\bibinfo {volume} {79}},\ \bibinfo {pages} {053821} (\bibinfo {year} {2009})}\BibitemShut {NoStop}%
\bibitem [{sup()}]{supple}%
  \BibitemOpen
  \href@noop {} {\bibinfo  {journal} {See {Supplementary Material} at (the link) for more details on { SI-1. Derivation of temporal Smith-Purcell radiation dispersion equation; SI-2. Simulation methods; SI-3. Energy band analysis for time grating; SI-4. Temporal Smith-Purcell radiation with high-energy electrons; SI-5. Comparison of temporal and spatial Smith-Purcell radiation; SI-6. Tuning radiation intensity through modulation amplitudes; SI-7. Temporal Smith-Purcell radiation with a temporal slab of finite thickness}}\ }\BibitemShut {NoStop}%
\bibitem [{\citenamefont {Pacheco-Pe{\~n}a}\ and\ \citenamefont {Engheta}(2020{\natexlab{c}})}]{pacheco2020effective}%
  \BibitemOpen
\bibfield  {journal} {  }\bibfield  {author} {\bibinfo {author} {\bibfnamefont {V.}~\bibnamefont {Pacheco-Pe{\~n}a}}\ and\ \bibinfo {author} {\bibfnamefont {N.}~\bibnamefont {Engheta}},\ }\href@noop {} {\bibfield  {journal} {\bibinfo  {journal} {Nanophotonics}\ }\textbf {\bibinfo {volume} {9}},\ \bibinfo {pages} {379} (\bibinfo {year} {2020}{\natexlab{c}})}\BibitemShut {NoStop}%
\bibitem [{\citenamefont {De~Abajo}(2010)}]{de2010optical}%
  \BibitemOpen
  \bibfield  {author} {\bibinfo {author} {\bibfnamefont {F.~G.}\ \bibnamefont {De~Abajo}},\ }\href@noop {} {\bibfield  {journal} {\bibinfo  {journal} {Reviews of Modern Physics}\ }\textbf {\bibinfo {volume} {82}},\ \bibinfo {pages} {209} (\bibinfo {year} {2010})}\BibitemShut {NoStop}%
\bibitem [{\citenamefont {Lu}\ \emph {et~al.}(2023)\citenamefont {Lu}, \citenamefont {Nussupbekov}, \citenamefont {Xiong}, \citenamefont {Ding}, \citenamefont {Png}, \citenamefont {Ooi}, \citenamefont {Teng}, \citenamefont {Wong}, \citenamefont {Chong},\ and\ \citenamefont {Wu}}]{lu2023smith}%
  \BibitemOpen
  \bibfield  {author} {\bibinfo {author} {\bibfnamefont {S.}~\bibnamefont {Lu}}, \bibinfo {author} {\bibfnamefont {A.}~\bibnamefont {Nussupbekov}}, \bibinfo {author} {\bibfnamefont {X.}~\bibnamefont {Xiong}}, \bibinfo {author} {\bibfnamefont {W.~J.}\ \bibnamefont {Ding}}, \bibinfo {author} {\bibfnamefont {C.~E.}\ \bibnamefont {Png}}, \bibinfo {author} {\bibfnamefont {Z.-E.}\ \bibnamefont {Ooi}}, \bibinfo {author} {\bibfnamefont {J.~H.}\ \bibnamefont {Teng}}, \bibinfo {author} {\bibfnamefont {L.~J.}\ \bibnamefont {Wong}}, \bibinfo {author} {\bibfnamefont {Y.}~\bibnamefont {Chong}}, \ and\ \bibinfo {author} {\bibfnamefont {L.}~\bibnamefont {Wu}},\ }\href@noop {} {\bibfield  {journal} {\bibinfo  {journal} {Laser \& Photonics Reviews}\ ,\ \bibinfo {pages} {2300002}} (\bibinfo {year} {2023})}\BibitemShut {NoStop}%
\bibitem [{\citenamefont {Xi}\ \emph {et~al.}(2009)\citenamefont {Xi}, \citenamefont {Chen}, \citenamefont {Jiang}, \citenamefont {Ran}, \citenamefont {Huangfu}, \citenamefont {Wu}, \citenamefont {Kong},\ and\ \citenamefont {Chen}}]{xi2009experimental}%
  \BibitemOpen
  \bibfield  {author} {\bibinfo {author} {\bibfnamefont {S.}~\bibnamefont {Xi}}, \bibinfo {author} {\bibfnamefont {H.}~\bibnamefont {Chen}}, \bibinfo {author} {\bibfnamefont {T.}~\bibnamefont {Jiang}}, \bibinfo {author} {\bibfnamefont {L.}~\bibnamefont {Ran}}, \bibinfo {author} {\bibfnamefont {J.}~\bibnamefont {Huangfu}}, \bibinfo {author} {\bibfnamefont {B.-I.}\ \bibnamefont {Wu}}, \bibinfo {author} {\bibfnamefont {J.~A.}\ \bibnamefont {Kong}}, \ and\ \bibinfo {author} {\bibfnamefont {M.}~\bibnamefont {Chen}},\ }\href@noop {} {\bibfield  {journal} {\bibinfo  {journal} {Physical Review Letters}\ }\textbf {\bibinfo {volume} {103}},\ \bibinfo {pages} {194801} (\bibinfo {year} {2009})}\BibitemShut {NoStop}%
\bibitem [{\citenamefont {Duan}\ \emph {et~al.}(2017)\citenamefont {Duan}, \citenamefont {Tang}, \citenamefont {Wang}, \citenamefont {Zhang}, \citenamefont {Chen}, \citenamefont {Chen},\ and\ \citenamefont {Gong}}]{duan2017observation}%
  \BibitemOpen
  \bibfield  {author} {\bibinfo {author} {\bibfnamefont {Z.}~\bibnamefont {Duan}}, \bibinfo {author} {\bibfnamefont {X.}~\bibnamefont {Tang}}, \bibinfo {author} {\bibfnamefont {Z.}~\bibnamefont {Wang}}, \bibinfo {author} {\bibfnamefont {Y.}~\bibnamefont {Zhang}}, \bibinfo {author} {\bibfnamefont {X.}~\bibnamefont {Chen}}, \bibinfo {author} {\bibfnamefont {M.}~\bibnamefont {Chen}}, \ and\ \bibinfo {author} {\bibfnamefont {Y.}~\bibnamefont {Gong}},\ }\href@noop {} {\bibfield  {journal} {\bibinfo  {journal} {Nature Communications}\ }\textbf {\bibinfo {volume} {8}},\ \bibinfo {pages} {14901} (\bibinfo {year} {2017})}\BibitemShut {NoStop}%
\bibitem [{\citenamefont {Chen}\ and\ \citenamefont {Chen}(2011)}]{chen2011flipping}%
  \BibitemOpen
  \bibfield  {author} {\bibinfo {author} {\bibfnamefont {H.}~\bibnamefont {Chen}}\ and\ \bibinfo {author} {\bibfnamefont {M.}~\bibnamefont {Chen}},\ }\href@noop {} {\bibfield  {journal} {\bibinfo  {journal} {Materials Today}\ }\textbf {\bibinfo {volume} {14}},\ \bibinfo {pages} {34} (\bibinfo {year} {2011})}\BibitemShut {NoStop}%
\bibitem [{\citenamefont {Liu}\ \emph {et~al.}(2017)\citenamefont {Liu}, \citenamefont {Xiao}, \citenamefont {Ye}, \citenamefont {Wang}, \citenamefont {Cui}, \citenamefont {Feng}, \citenamefont {Zhang},\ and\ \citenamefont {Huang}}]{liu2017integrated}%
  \BibitemOpen
  \bibfield  {author} {\bibinfo {author} {\bibfnamefont {F.}~\bibnamefont {Liu}}, \bibinfo {author} {\bibfnamefont {L.}~\bibnamefont {Xiao}}, \bibinfo {author} {\bibfnamefont {Y.}~\bibnamefont {Ye}}, \bibinfo {author} {\bibfnamefont {M.}~\bibnamefont {Wang}}, \bibinfo {author} {\bibfnamefont {K.}~\bibnamefont {Cui}}, \bibinfo {author} {\bibfnamefont {X.}~\bibnamefont {Feng}}, \bibinfo {author} {\bibfnamefont {W.}~\bibnamefont {Zhang}}, \ and\ \bibinfo {author} {\bibfnamefont {Y.}~\bibnamefont {Huang}},\ }\href@noop {} {\bibfield  {journal} {\bibinfo  {journal} {Nature Photonics}\ }\textbf {\bibinfo {volume} {11}},\ \bibinfo {pages} {289} (\bibinfo {year} {2017})}\BibitemShut {NoStop}%
\bibitem [{\citenamefont {Gaxiola-Luna}\ and\ \citenamefont {Halevi}(2021)}]{gaxiola2021temporal}%
  \BibitemOpen
  \bibfield  {author} {\bibinfo {author} {\bibfnamefont {J.~G.}\ \bibnamefont {Gaxiola-Luna}}\ and\ \bibinfo {author} {\bibfnamefont {P.}~\bibnamefont {Halevi}},\ }\href@noop {} {\bibfield  {journal} {\bibinfo  {journal} {Physical Review B}\ }\textbf {\bibinfo {volume} {103}},\ \bibinfo {pages} {144306} (\bibinfo {year} {2021})}\BibitemShut {NoStop}%
\bibitem [{\citenamefont {Gaxiola-Luna}\ and\ \citenamefont {Halevi}(2023)}]{gaxiola2023growing}%
  \BibitemOpen
  \bibfield  {author} {\bibinfo {author} {\bibfnamefont {J.}~\bibnamefont {Gaxiola-Luna}}\ and\ \bibinfo {author} {\bibfnamefont {P.}~\bibnamefont {Halevi}},\ }\href@noop {} {\bibfield  {journal} {\bibinfo  {journal} {Applied Physics Letters}\ }\textbf {\bibinfo {volume} {122}} (\bibinfo {year} {2023})}\BibitemShut {NoStop}%
\bibitem [{\citenamefont {Ye}\ \emph {et~al.}(2019)\citenamefont {Ye}, \citenamefont {Liu}, \citenamefont {Wang}, \citenamefont {Tai}, \citenamefont {Cui}, \citenamefont {Feng}, \citenamefont {Zhang},\ and\ \citenamefont {Huang}}]{ye2019deep}%
  \BibitemOpen
  \bibfield  {author} {\bibinfo {author} {\bibfnamefont {Y.}~\bibnamefont {Ye}}, \bibinfo {author} {\bibfnamefont {F.}~\bibnamefont {Liu}}, \bibinfo {author} {\bibfnamefont {M.}~\bibnamefont {Wang}}, \bibinfo {author} {\bibfnamefont {L.}~\bibnamefont {Tai}}, \bibinfo {author} {\bibfnamefont {K.}~\bibnamefont {Cui}}, \bibinfo {author} {\bibfnamefont {X.}~\bibnamefont {Feng}}, \bibinfo {author} {\bibfnamefont {W.}~\bibnamefont {Zhang}}, \ and\ \bibinfo {author} {\bibfnamefont {Y.}~\bibnamefont {Huang}},\ }\href@noop {} {\bibfield  {journal} {\bibinfo  {journal} {Optica}\ }\textbf {\bibinfo {volume} {6}},\ \bibinfo {pages} {592} (\bibinfo {year} {2019})}\BibitemShut {NoStop}%
\bibitem [{\citenamefont {Kong}(1975)}]{kong1975theory}%
  \BibitemOpen
  \bibfield  {author} {\bibinfo {author} {\bibfnamefont {J.~A.}\ \bibnamefont {Kong}},\ }\href@noop {} {\bibfield  {journal} {\bibinfo  {journal} {New York}\ } (\bibinfo {year} {1975})}\BibitemShut {NoStop}%
\bibitem [{\citenamefont {Zhu}\ \emph {et~al.}(2019)\citenamefont {Zhu}, \citenamefont {Du}, \citenamefont {Li}, \citenamefont {Bao},\ and\ \citenamefont {Liu}}]{zhu2019free}%
  \BibitemOpen
  \bibfield  {author} {\bibinfo {author} {\bibfnamefont {J.-F.}\ \bibnamefont {Zhu}}, \bibinfo {author} {\bibfnamefont {C.-H.}\ \bibnamefont {Du}}, \bibinfo {author} {\bibfnamefont {F.-H.}\ \bibnamefont {Li}}, \bibinfo {author} {\bibfnamefont {L.-Y.}\ \bibnamefont {Bao}}, \ and\ \bibinfo {author} {\bibfnamefont {P.-K.}\ \bibnamefont {Liu}},\ }\href@noop {} {\bibfield  {journal} {\bibinfo  {journal} {IEEE Access}\ }\textbf {\bibinfo {volume} {7}},\ \bibinfo {pages} {181184} (\bibinfo {year} {2019})}\BibitemShut {NoStop}%
\bibitem [{\citenamefont {Zhang}\ \emph {et~al.}(2023{\natexlab{b}})\citenamefont {Zhang}, \citenamefont {Zeng}, \citenamefont {Tian},\ and\ \citenamefont {Li}}]{zhang2023coherent}%
  \BibitemOpen
  \bibfield  {author} {\bibinfo {author} {\bibfnamefont {D.}~\bibnamefont {Zhang}}, \bibinfo {author} {\bibfnamefont {Y.}~\bibnamefont {Zeng}}, \bibinfo {author} {\bibfnamefont {Y.}~\bibnamefont {Tian}}, \ and\ \bibinfo {author} {\bibfnamefont {R.}~\bibnamefont {Li}},\ }\href@noop {} {\bibfield  {journal} {\bibinfo  {journal} {Photonics Insights}\ }\textbf {\bibinfo {volume} {2}},\ \bibinfo {pages} {R07} (\bibinfo {year} {2023}{\natexlab{b}})}\BibitemShut {NoStop}%
\bibitem [{\citenamefont {Chlouba}\ \emph {et~al.}(2023)\citenamefont {Chlouba}, \citenamefont {Shiloh}, \citenamefont {Kraus}, \citenamefont {Br{\"u}ckner}, \citenamefont {Litzel},\ and\ \citenamefont {Hommelhoff}}]{chlouba2023coherent}%
  \BibitemOpen
  \bibfield  {author} {\bibinfo {author} {\bibfnamefont {T.}~\bibnamefont {Chlouba}}, \bibinfo {author} {\bibfnamefont {R.}~\bibnamefont {Shiloh}}, \bibinfo {author} {\bibfnamefont {S.}~\bibnamefont {Kraus}}, \bibinfo {author} {\bibfnamefont {L.}~\bibnamefont {Br{\"u}ckner}}, \bibinfo {author} {\bibfnamefont {J.}~\bibnamefont {Litzel}}, \ and\ \bibinfo {author} {\bibfnamefont {P.}~\bibnamefont {Hommelhoff}},\ }\href@noop {} {\bibfield  {journal} {\bibinfo  {journal} {Nature}\ }\textbf {\bibinfo {volume} {622}},\ \bibinfo {pages} {476} (\bibinfo {year} {2023})}\BibitemShut {NoStop}%
\bibitem [{\citenamefont {Zhang}\ \emph {et~al.}(2022)\citenamefont {Zhang}, \citenamefont {Zeng}, \citenamefont {Bai}, \citenamefont {Li}, \citenamefont {Tian},\ and\ \citenamefont {Li}}]{zhang2022coherent}%
  \BibitemOpen
  \bibfield  {author} {\bibinfo {author} {\bibfnamefont {D.}~\bibnamefont {Zhang}}, \bibinfo {author} {\bibfnamefont {Y.}~\bibnamefont {Zeng}}, \bibinfo {author} {\bibfnamefont {Y.}~\bibnamefont {Bai}}, \bibinfo {author} {\bibfnamefont {Z.}~\bibnamefont {Li}}, \bibinfo {author} {\bibfnamefont {Y.}~\bibnamefont {Tian}}, \ and\ \bibinfo {author} {\bibfnamefont {R.}~\bibnamefont {Li}},\ }\href@noop {} {\bibfield  {journal} {\bibinfo  {journal} {Nature}\ }\textbf {\bibinfo {volume} {611}},\ \bibinfo {pages} {55} (\bibinfo {year} {2022})}\BibitemShut {NoStop}%
\bibitem [{\citenamefont {Sharabi}\ \emph {et~al.}(2021)\citenamefont {Sharabi}, \citenamefont {Lustig},\ and\ \citenamefont {Segev}}]{sharabi2021disordered}%
  \BibitemOpen
  \bibfield  {author} {\bibinfo {author} {\bibfnamefont {Y.}~\bibnamefont {Sharabi}}, \bibinfo {author} {\bibfnamefont {E.}~\bibnamefont {Lustig}}, \ and\ \bibinfo {author} {\bibfnamefont {M.}~\bibnamefont {Segev}},\ }\href@noop {} {\bibfield  {journal} {\bibinfo  {journal} {Physical Review Letters}\ }\textbf {\bibinfo {volume} {126}},\ \bibinfo {pages} {163902} (\bibinfo {year} {2021})}\BibitemShut {NoStop}%
\bibitem [{\citenamefont {Phare}\ \emph {et~al.}(2015)\citenamefont {Phare}, \citenamefont {Daniel~Lee}, \citenamefont {Cardenas},\ and\ \citenamefont {Lipson}}]{phare2015graphene}%
  \BibitemOpen
  \bibfield  {author} {\bibinfo {author} {\bibfnamefont {C.~T.}\ \bibnamefont {Phare}}, \bibinfo {author} {\bibfnamefont {Y.-H.}\ \bibnamefont {Daniel~Lee}}, \bibinfo {author} {\bibfnamefont {J.}~\bibnamefont {Cardenas}}, \ and\ \bibinfo {author} {\bibfnamefont {M.}~\bibnamefont {Lipson}},\ }\href@noop {} {\bibfield  {journal} {\bibinfo  {journal} {Nature Photonics}\ }\textbf {\bibinfo {volume} {9}},\ \bibinfo {pages} {511} (\bibinfo {year} {2015})}\BibitemShut {NoStop}%
\end{thebibliography}%

\end{document}